# Effective Classification of MicroRNA Precursors Using Combinatorial Feature Mining and AdaBoost Algorithms

Ling Zhong and Jason T. L. Wang

Bioinformatics Program and Department of Computer Science, New Jersey Institute of Technology, Newark, New Jersey 07102, USA

**Abstract**

MicroRNAs (miRNAs) are non-coding RNAs with approximately 22 nucleotides (nt) that are derived from precursor molecules. These precursor molecules or pre-miRNAs often fold into stem-loop hairpin structures. However, a large number of sequences with pre-miRNA-like hairpins can be found in genomes. It is a challenge to distinguish the real pre-miRNAs from other hairpin sequences with similar stem-loops (referred to as pseudo pre-miRNAs). Several computational methods have been developed to tackle this challenge. In this paper we propose a new method, called MirID, for identifying and classifying microRNA precursors. We collect 74 features from the sequences and secondary structures of pre-miRNAs; some of these features are taken from our previous studies on non-coding RNA prediction while others were suggested in the literature. We develop a combinatorial feature mining algorithm to identify suitable feature sets. These feature sets are then used to train support vector machines to obtain classification models, based on which classifier ensemble is constructed. Finally we use an AdaBoost algorithm to further enhance the accuracy of the classifier ensemble. Experimental results on a variety of species demonstrate the good performance of the proposed method, and its superiority over existing tools.

# Introduction

MicroRNAs (miRNAs) are non-coding RNAs (ncRNAs) of approximately 22 nucleotides that are known to regulate post-transcriptional expression of protein-coding genes (Bartel, 2004; Bindra et al., 2010). Lee et al. (1993) first reported that in *C. elegans*, lin-4 regulates the translation of lin-14 mRNA via an antisense RNA-RNA interaction. Since then, many roles of miRNAs have been discovered, including for example the control of left/right neuronal asymmetry in *C. elegans* (Johnston and Hobert, 2003), cell proliferation in *Drosophila* (Brennecke et al., 2003), and the regulation of flowering time in plants (Aukerman and Sakai, 2003). Their role in cancer development has also been reported (Mack, 2007). However, these roles are only a small portion of total miRNA functions (Bushati and Cohen, 2007). As a consequence, exploring miRNAs and their functions continues to be a highly active area of research.

MiRNAs are derived from pre-miRNAs that often fold into stem-loop hairpin structures. These characteristic stem-loop structures are highly conserved in different species (Lai et al., 2003). One challenging research problem is to distinguish pre-miRNAs from other sequences with similar stem-loop structures (referred to as pseudo pre-miRNAs). Many computational methods have been developed to tackle this challenge. A common approach is to transform the classification of real and pseudo pre-miRNAs to a feature selection problem.

Lim et al. (2003) reported some characteristic features in phylogenetically conserved stem loop pre-miRNAs. Lai et al. (2003) considered hairpin structures predicted by mfold (Zuker, 2003) as well as the nucleotide divergence of pre-miRNAs. Xue et al. (2005) decomposed stem-loop hairpin structures into local structure-sequence features, and used these features in combination with a support vector machine to classify pre-miRNAs. Bentwich et al. (2005) proposed a scoring function for pre-miRNAs with thermodynamic stability and certain structural features, which capture the global properties of the hairpin structures in the pre-miRNAs. Ng and Mishra (2007) employed a Gaussian radial basis function kernel as a similarity measure for 29 global and intrinsic hairpin folding attributes, and characterized pre-miRNAs based on their dinucleotide subsequences, hairpin folding, non-linear statistical thermodynamics and topology. Huang et al. (2007) evaluated features valuable for pre-miRNA classification, such as the local secondary structure differences of the stem regions of real pre-miRNA and pseudo pre-miRNA hairpins, and established correlations between different types of mutations and the secondary structures of real pre-miRNAs. More recently, Zhao et al. (2010) considered structure-sequence features and minimum free energy of RNA secondary structure, along with the double helix structure with free nucleotides and base-pairing features. In general, the quality of selected features directly affects the classification accuracy achieved by a method.

In this paper, we present a novel combinatorial feature mining method for pre-miRNA classification. Our method, named MirID, identifies and classifies an input RNA sequence as a pre-miRNA or not. MirID considers different combinations of features extracted from pre-miRNAs. For each combination (or each set of features), we create a support vector machine (SVM) model (Cortes and Vapnik, 1995; Fan et al., 2005) based on that feature set. SVM models whose accuracies are above a user-determined threshold are then used to build a classifier ensemble. This classifier ensemble will be refined through several iterations until its accuracy cannot be enhanced further. We next construct new feature sets based on the best feature sets obtained so far by performing pairwise merge and split operations on the best feature sets. Then we repeat the above procedure iteratively by building a SVM model based on each new feature set, constructing a classifier ensemble from the SVM models whose accuracies are above the newly computed threshold, and refining the ensemble until it can not be improved further. Finally we output the best classifier ensemble obtained through this iterative procedure. To further enhance the accuracy of the classifier ensemble, we apply a boosting algorithm to the ensemble to obtain a strong classifier, which is used for pre-miRNA classification.

**Materials and Methods**

*Datasets*

We collected real pre-miRNAs and pseudo pre-miRNAs from twenty one species, some of which were studied previously while others have not been explored. These RNA sequences were evenly divided into training data and test data. Table 1 presents a summary of the data. The first column of Table 1 shows a species or organism name. The second column of Table 1 shows the number of training sequences followed by the number of test sequences with respect to the organism's real pre-miRNAs. The third column of Table 1 shows the number of training sequences followed by the number of test sequences with respect to the organism's pseudo pre-miRNAs. As an example, referring to *Arabidopsis thaliana* in Table 1, its training set contains 66 real pre-miRNAs and 923 pseudo pre-miRNAs; its test set contains 67 real pre-miRNAs and 924 pseudo pre-miRNAs.

The real pre-miRNAs were downloaded from miRBase available at http://www.mirbase.org/ (Kozomara and Griffiths-Jones, 2011). We used RNAfold (Hofacker, 2003) to predict the secondary structures of all the RNA sequences. The lengths of the real pre-miRNAs in the final dataset ranged from 60 to 120 nt. The pseudo pre-miRNAs used in this study were collected from GenBank (http://www.ncbi.nlm.nih.gov/genbank/). As in (Xue et al., 2005), we searched for the protein-coding regions of the genome sequences of the twenty one species in Table 1, and divided the regions into short sequences. The lengths of these short sequences were randomly generated, ranging from 60 to 120 nt. The pseudo pre-miRNAs were chosen from these short sequences. The criteria used in choosing the pseudo pre-miRNAs are: (i) they have a stem-loop hairpin structure, (ii) they contain at least 18 base pairs, including Watson-Crick and wobble base pairs, on the stem region of the stem-loop structure, and (iii) their secondary structure has a maximum of -15 kcal/mol free energy without multiple hairpin loops (Kozomara and Griffiths-Jones, 2011). These criteria ensure that the secondary structures of the pseudo pre-miRNAs are similar to those of the real pre-miRNAs.

*Feature pool*

In designing our pre-miRNA classification method, we examined multiple features extracted from a pre-miRNA sequence and its secondary structure. Some of these features were taken from our previous studies on ncRNA prediction (Griesmer et al., 2011; Wang and Wu, 2006) while others were suggested in the literature (Sewer et al., 2005; Xue et al., 2005; Zheng et al., 2006). These features included the sequence length, the number of base pairs, GC content, the number of nucleotides contained in the hairpin loop (i.e., the loop size), the free energy of the sequence's secondary structure obtained from RNAfold (Hofacker, 2003), the number of bulge loops, and the size of the largest bulge loop in the secondary structure.

In addition, we considered the features described in (Zheng et al., 2006). These features included the difference of the lengths of the two tails in the secondary structure where a tail represented the strand of unpaired bases in the 5′ or 3′ end of the structure, the number of tails, and the length of the larger tail. Besides, several combined features were considered. They included the ratio between the number of base pairs and the sequence length, the length difference of two tails plus the larger tail length, the size of the hairpin loop plus the larger tail length, the size of the hairpin loop plus the largest bulge size, the ratio between the larger tail length and the sequence length, the ratio between the size of the hairpin loop and the sequence length, the ratio between the largest bulge size and the sequence length, the ratio between the largest bulge size and the number of base pairs, the normalized free energy (Spirollari et al., 2009), which is the minimum free energy of the sequence's secondary structure divided by the sequence length, and the ratio between the normalized free energy and the GC content.

The next set of features included the triplets of structure-sequence elements described in (Xue et al., 2005). Here we used the dot-bracket notation (Hofacker, 2003) to represent an RNA secondary structure. Figure 1 shows the sequence and structure of a hypothetical pre-miRNA and its dot-bracket notation. A triplet is composed of three contiguous structure elements (bases or base pairs) (Liu et al., 2005) that correspond to three contiguous nucleotides along with the middle nucleotide. For example, consider the first three dots (bases) and their corresponding nucleotides AAA in Figure 1. The middle nucleotide is A. Thus, the structure-sequence elements "A..." constitute a triplet. As another example, consider the first three brackets (base pairs) and their corresponding nucleotides UUG in Figure 1. The middle nucleotide is U. Thus, the structure-sequence elements "U(((" constitute a triplet. There are 32 triplets, and hence 32 such features in total.

Finally we considered the symmetric and asymmetric loops defined in (Sewer et al., 2005). We refer to the portion of the sequence from the 5′ end to the hairpin loop as the left arm, and the portion of the sequence from the hairpin loop to the 3′ end as the right arm. In a symmetric (internal) loop, the number of nucleotides in the left arm equals the number of nucleotides in the right arm. In an asymmetric (internal) loop, the number of nucleotides in the left arm is different from the number of nucleotides in the right arm. Features related to these loops included the size of each loop, the average size of the loops, and the average distance between the loops. Other features included the proportion of A/C/G/U in the stem, and the proportion of A-U/C-G/G-U base pairs in the stem. Totally, there are 74 features in the feature pool.

*Combinatorial feature mining*

MirID adopts a novel feature mining algorithm for pre-miRNA classification. Initially the algorithm randomly generates $N$ feature sets from the feature pool. (The default value of $N$ used in this study is 100.) Each feature set contains between 1 and 150 features, randomly chosen with replacement from the feature pool. Some features may repeatedly occur in a feature set; thus a bagging approach is used here (Breiman, 1996).

Duplicate features have more weights than the other features in the feature set. The numbers 1 and 150 are chosen, to ensure that there are enough feature sets containing duplicate features. We then build a SVM model based on each feature set using training sequences, and apply the classification model to test sequences to calculate the accuracy of the model. The SVM used in this study is the LIBSVM package downloaded from http://www.csie.ntu.edu.tw/~cjlin/libsvm/ (Fan et al., 2005). We use the polynomial kernel provided in the LIBSVM package. The polynomial kernel achieves the best performance among all kernel functions included in the package.

Then, we remove the SVM models whose accuracies are less than a user-determined threshold $t$. (The default value of $t$ used in this study is 0.8.) The feature sets used to build those removed SVM models are also eliminated from further consideration. We construct a classifier ensemble from the remaining SVM models. The ensemble works by taking the majority vote from the individual SVM models used to build the classifier ensemble. This ensemble will be refined through several iterations until its accuracy cannot be enhanced further. In each iteration, the user-determined threshold $t$ is incremented by a *step* value, so that more accurate SVM models are used to construct a (hopefully) better classifier ensemble in the next iteration. (The default value of *step* used in this study is 0.005.)

It is likely that different combinations of remaining features may yield an even better classifier. Our algorithm then performs pairwise *merge* and *split* operations on the set $S_b$ of feature sets used to build the best classifier ensemble obtained so far. In doing so, MirID takes four steps: (1) picks each pair of feature sets $s_1$ and $s_2$ in $S_b$; (2) merges $s_1$ and $s_2$ into a single feature set $s_3$ with, say $p$, features; (3) randomly generates a number $q$, $q < p$; (4) randomly assigns $q$ features in $s_3$ into a set $s'_1$ and assigns the remaining $p - q$ features into another set $s'_2$. Thus, these four steps take two feature sets $s_1$ and $s_2$ in $S_b$ as input and produce two new feature sets $s'_1$ and $s'_2$ as output. Figure 2 illustrates how the merge and split operations work on two feature sets.

These pairwise merge and split operations are applied to the feature sets used to build the best classifier ensemble obtained so far, to generate new feature sets. The new feature sets are then used to build new SVM models. Accurate new SVM models, whose accuracies are greater than or equal to the newly computed threshold $t$, are then used to build a new classifier ensemble. This procedure is repeated several times to obtain a best classifier ensemble. Figure 3 summarizes our feature mining algorithm, whose output is the best classifier ensemble along with the component SVM models (feature sets) used to build the ensemble. Notice that in the feature mining algorithm in Figure 3, it is possible that, after removing SVM models/feature sets with accuracy $< t$, there is no remaining feature set, and hence $S_r$ becomes empty. Under this circumstance, the classifier ensemble constructed based on $S_r$ is empty, and the accuracy of the classifier ensemble is 0.

*Boosting*

The performance of a classification algorithm can be further enhanced through boosting. We apply AdaBoost (Bindewald and Shapiro, 2006; Freund and Schapire, 1997; Schapire, 1999) to the classifier ensemble produced by our feature mining algorithm. Specifically, we treat the classifier ensemble as a weak classifier and continue refining it into a strong classifier through an iterative procedure. Let $X$ be a set of sequences $x_1$, $x_2, \ldots, x_m$ where $x_i$, $1 \leq i \leq m$, is associated with a label $y_i$ such that

$$y_i = \begin{cases} +1 & \text{if } x_i \text{ is a real pre-miRNA} \\ -1 & \text{if } x_i \text{ is a pseudo pre-miRNA} \end{cases}$$

The AdaBoost algorithm works with $K$ iterations. (The default value of $K$ used in this study is 20.) In iteration $k$, $1 \leq k \leq K$, the algorithm updates a weight function $W_k$ as explained below, which will be used in selecting training sequences in iteration $k + 1$. Initially, every sequence has an equal weight, i.e. $W_0(x_i) = 1/m$, $1 \leq i \leq m$. In iteration $k$, the algorithm samples 1/3 sequences with replacement from $X$ based on the weight function $W_{k-1}$ to form a training set $X_k$. The set $X_k$ is then used to train a weak classifier $H_k$, which classifies each sequence $x_i$ as either a real pre-miRNA or a pseudo pre-miRNA. That is,

$$H_k(x_i) = \begin{cases} +1 & H_k \text{ classifies } x_i \text{ as a real pre-miRNA} \\ -1 & H_k \text{ classifies } x_i \text{ as a pseudo pre-miRNA} \end{cases}$$

Let $E_k = \{x_i | H_k(x_i) \neq y_i\}$. The error rate $\varepsilon_k$ of $H_k$ is:

$$\varepsilon_k = \sum_{x_i \in E_k} W_{k-1}(x_i) \tag{1}$$

Let

$$\alpha_k = \frac{1}{2} \ln\left(\frac{1-\varepsilon_k}{\varepsilon_k}\right) \tag{2}$$

The algorithm updates $W_k$ for each sequence $x_i$, $1 \leq i \leq m$, as follows:

$$W_k(x_i) = \begin{cases} \dfrac{W_{k-1}(x_i)}{Z_k} \times e^{-\alpha_k} & \text{if } H_k(x_i) = y_i \\ \dfrac{W_{k-1}(x_i)}{Z_k} \times e^{\alpha_k} & \text{if } H_k(x_i) \neq y_i \end{cases}$$

$$= \frac{W_{k-1}(x_i) \exp(-\alpha_k y_i H_k(x_i))}{Z_k} \tag{3}$$

where $Z_k$ is a normalization factor chosen such that $W_k$ is normally distributed. Thus, the sequences causing classification errors in iteration $k$ will have a greater probability of being selected as training sequences for constructing the weak classifier $H_{k+1}$ in iteration $k+1$. Using this technique, each weak classifier should have greater accuracy than its predecessor. The final, strong classifier $H$ combines the vote of each individual weak classifier $H_k$, $1 \leq k \leq K$, where the weight of each weak classifier's vote is a function of its accuracy. Specifically, for an unlabeled test sequence $x$, $H(x)$ is calculated as follows:

$$H(x) = sign\left(\sum_{k=1}^{K} \alpha_k H_k(x)\right) \qquad (4)$$

The function *sign* indicates that if the sum inside the parentheses is greater than or equal to zero, then $H$ classifies $x$ as positive (i.e. a real pre-miRNA); otherwise $H$ classifies $x$ as negative (i.e. a pseudo pre-miRNA).

## Results

*Performance analysis of the MirID method*

We carried out a series of experiments to evaluate the proposed MirID method. All the experiments were performed on a 2 GHz Pentium 4 PC having a memory of 2G bytes. The operating system was Cygwin on Windows XP and the algorithms were implemented in Perl. To understand the effect of boosting, we also considered using the combinatorial feature mining algorithm alone to classify pre-miRNAs, and referred to it as the CFM method. The performance measure used here is accuracy, defined as follows. A method is said to classify a test sequence correctly if the sequence is a real pre-miRNA (pseudo pre-miRNA, respectively) and the method indicates that the sequence is indeed a real pre-miRNA (pseudo pre-miRNA, respectively). A method is said to classify a test sequence incorrectly if the sequence is a real pre-miRNA (pseudo pre-miRNA, respectively) but the method mistakenly indicates that the sequence is a pseudo pre-miRNA (real pre-miRNA, respectively). For each species, the accuracy of a method is defined as the number of correctly classified test sequences of that species divided by the total number of test sequences of that species.

We first evaluated how the number of initial feature sets, $N$, affects the performance of CFM and MirID. As $N$ increases, more feature sets are generated initially. This allows the feature mining algorithm to construct a classifier ensemble using more diverse feature sets, and hence the accuracy of the classifier ensemble increases. On the other hand, as $N$ increases, the inner loop in Figure 3 is run more times; as a consequence, the running time increases. MirID requires more time than CFM, due to the extra time spent in boosting. MirID in general is more accurate than CFM, indicating the benefit of including the boosting algorithm.

We next evaluated how the threshold, $t$, used in the feature mining algorithm affects the performance of CFM and MirID. When $t$ is very large (e.g. $t > 0.95$), the accuracies of the methods drop sharply. This happens because the accuracies of most SVM models are less than 0.95, and hence these SVM models are eliminated from further consideration early in the feature mining algorithm, cf. Figure 3. When $t$ approaches 1, it is likely that the set $S_b$ returned by the feature mining algorithm is an empty set, and therefore the classifier ensemble constructed based on $S_b$ is also empty, yielding an accuracy of 0. As $t$ increases, fewer feature sets qualify and the set $S_r$ is smaller. As a result, the inner loop in Figure 3 is executed fewer times, and hence the running time decreases.

Then we evaluated how the value, *step*, used to increment the threshold *t* in each iteration of the inner loop in Figure 3 affects the performance of CFM and MirID. With the default values of *N* and *t* used in this study, the feature mining algorithm is able to produce a classifier ensemble with high accuracy. The value of *step* has little impact on the accuracies of the proposed methods. However, as *step* increases, fewer iterations of the inner loop in Figure 3 are executed, and as a consequence, the running time decreases.

We also conducted experiments to test different numbers of iterations, *K*, in the boosting algorithm. It was found that when *K* is sufficiently large (e.g. $K \geq 20$), the behavior of the boosting algorithm becomes stable, with the accuracy approaching 1. On the other hand, when *K* is large, more running time will be needed.

Finally we compared CFM and MirID with two closely related methods, PMirP (Zhao et al., 2010) and TripletSVM (Xue et al., 2005). Like our methods, both PMirP and TripletSVM were implemented using support vector machines. PMirP adopted a hybrid coding scheme, combining features such as free bases, base pairs, minimum free energy of secondary structure, among others. TripletSVM used triplets of structure-sequence elements, which also were included in our feature pool. Table 2 shows the accuracies of the four methods on twelve species taken from Table 1. These twelve species were used to pre-train PMirP and TripletSVM, and available from the tools. For each species, the highest accuracy yielded by a tool is in bold. It can be seen from Table 2 that MirID is better than or as good as the existing tools on all the species except *Gallus gallus* and *Oryza sativa.* For *Gallus gallus* and *Oryza sativa,* PMirP achieves higher accuracies.

*Software Tool*

We have implemented MirID using Perl into a software tool, available from the authors upon request. The software tool accepts a test sequence as input and classifies the test sequence as a pre-miRNA or not. We pre-train our software tool using the training sequences given in Table 1. In addition to the twelve species available from the PMirP and TripletSVM web servers (Xue et al., 2005; Zhao et al., 2010), we pre-train our software tool using nine additional species (shown in Table 1 but not in Table 2). Our tool achieves high accuracies on these nine species, as shown in Table 3. (The PMirP and TripletSVM web servers were not pre-trained on these nine species, and hence we only show the results for CFM and MirID here.) MirID is more accurate than CFM, due to the boosting algorithm.

Table 4 shows, for each species in Table 1, the number of feature sets produced by our feature mining algorithm. Table 5 shows the CPU time (in seconds) spent in pre-training the MirID tool. The training time depends on the number of feature sets, the number of features in each feature set, the number of iterations used by the feature mining algorithm, and the number of iterations used in the boosting algorithm. Notice that this training is done once, and no more training is needed on the test data. It takes less than a second to classify an unlabeled test sequence.

## Conclusions

In this paper we present a new method (MirID) for pre-miRNA classification. Empirical results showed that MirID outperforms two closely related methods, PMirP and TripletSVM, on the majority of species tested in the experiments. Since all the three methods were implemented using support vector machines with similar features, we conclude that the superiority of our method is due to its feature mining and boosting algorithms. In the future, we plan to extend these algorithms for classifying and predicting other RNA structures; see, e.g. (Laing et al., 2012).

TABLE 1. Summary of Datasets

| Species | Real pre-miRNA | Pseudo pre-miRNA |
|---|---|---|
| *Arabidopsis thaliana* | 66, 67 | 923, 924 |
| *Caenorhabditis briggsae* | 66, 67 | 437, 438 |
| *Caenorhabditis elegans* | 84, 85 | 595, 596 |
| *Canis familiaris* | 161, 161 | 904, 905 |
| *Ciona intestinalis* | 160, 160 | 733, 734 |
| *Danio rerio* | 170, 170 | 1071, 1072 |
| *Drosophila melanogaster* | 81, 82 | 694, 694 |
| *Drosophila pseudoobscura* | 98, 99 | 495, 495 |
| *Epstein barr virus* | 12, 13 | 119, 119 |
| *Gallus gallus* | 241, 241 | 1186, 1186 |
| *Homo sapiens* | 504, 504 | 1999, 2000 |
| *Macaca mulatta* | 222, 223 | 1086, 1086 |
| *Medicago truncatula* | 111, 111 | 116, 116 |
| *Mus musculus* | 315, 315 | 2019, 2019 |
| *Oryza sativa* | 172, 172 | 522, 523 |
| *Physcomitrella patens* | 73, 74 | 703, 704 |
| *Populus trichocarpa* | 94, 95 | 809, 810 |
| *Pristionchus pacificus* | 60, 61 | 58, 58 |
| *Rattus norvegicus* | 193, 193 | 1238, 1238 |
| *Schmidtea mediterranea* | 72, 73 | 201, 202 |
| *Taeniopygia guttata* | 94, 95 | 483, 483 |

TABLE 2. Comparison of Four Pre-miRNA Classification Methods

| Species | TripletSVM | PMirP | CFM | MirID |
|---|---|---|---|---|
| *Arabidopsis thaliana* | 0.92 | 0.96 | 0.99 | **1.00** |
| *Caenorhabditis briggsae* | 0.96 | 0.97 | 0.98 | **1.00** |
| *Caenorhabditis elegans* | 0.86 | 0.86 | 0.97 | **0.98** |
| *Danio rerio* | 0.67 | 0.83 | 0.98 | **0.99** |
| *Drosophila melanogaster* | 0.92 | 0.96 | 0.97 | **0.99** |
| *Drosophila pseudoobscura* | 0.90 | 0.92 | 0.98 | **1.00** |
| *Epstein barr virus* | **1.00** | 0.80 | 0.98 | **1.00** |
| *Gallus gallus* | 0.85 | **1.00** | 0.96 | 0.96 |
| *Homo sapiens* | 0.93 | **0.95** | 0.93 | **0.95** |
| *Mus musculus* | 0.94 | 0.94 | 0.95 | **0.97** |
| *Oryza sativa* | 0.95 | **1.00** | 0.97 | 0.99 |
| *Rattus norvegicus* | 0.80 | 0.92 | 0.97 | **0.98** |

TABLE 3. Accuracies of CFM and MirID on Nine Additional Species

| Species | CFM | MirID |
|---|---|---|
| *Canis familiaris* | 0.97 | **1.00** |
| *Ciona intestinalis* | 0.94 | **1.00** |
| *Macaca mulatta* | **0.96** | **0.96** |
| *Medicago truncatula* | 0.95 | **1.00** |
| *Physcomitrella patens* | **1.00** | **1.00** |
| *Populus trichocarpa* | 0.97 | **0.99** |
| *Pristionchus pacificus* | 0.96 | **1.00** |
| *Schmidtea mediterranea* | 0.95 | **0.99** |
| *Taeniopygia guttata* | 0.95 | **0.99** |

TABLE 4. Number of Feature Sets for Each Species in MirID

| Species | Number of feature sets |
|---|---|
| *Arabidopsis thaliana* | 1 |
| *Caenorhabditis briggsae* | 6 |
| *Caenorhabditis elegans* | 1 |
| *Canis familiaris* | 1 |
| *Ciona intestinalis* | 7 |
| *Danio rerio* | 11 |
| *Drosophila melanogaster* | 3 |
| *Drosophila pseudoobscura* | 4 |
| *Epstein barr virus* | 5 |
| *Gallus gallus* | 3 |
| *Homo sapiens* | 1 |
| *Macaca mulatta* | 1 |
| *Medicago truncatula* | 1 |
| *Mus musculus* | 3 |
| *Oryza sativa* | 3 |
| *Physcomitrella patens* | 1 |
| *Populus trichocarpa* | 1 |
| *Pristionchus pacificus* | 1 |
| *Rattus norvegicus* | 10 |
| *Schmidtea mediterranea* | 32 |
| *Taeniopygia guttata* | 5 |

TABLE 5. Training Time for Each Species in MirID

| Species | Training time |
|---|---|
| *Arabidopsis thaliana* | 80 |
| *Caenorhabditis briggsae* | 348 |
| *Caenorhabditis elegans* | 103 |
| *Canis familiaris* | 153 |
| *Ciona intestinalis* | 269 |
| *Danio rerio* | 1272 |
| *Drosophila melanogaster* | 199 |
| *Drosophila pseudoobscura* | 196 |
| *Epstein barr virus* | 113 |
| *Gallus gallus* | 274 |
| *Homo sapiens* | 1530 |
| *Macaca mulatta* | 243 |
| *Medicago truncatula* | 104 |
| *Mus musculus* | 786 |
| *Oryza sativa* | 214 |
| *Physcomitrella patens* | 90 |
| *Populus trichocarpa* | 138 |
| *Pristionchus pacificus* | 63 |
| *Rattus norvegicus* | 349 |
| *Schmidtea mediterranea* | 478 |
| *Taeniopygia guttata* | 156 |

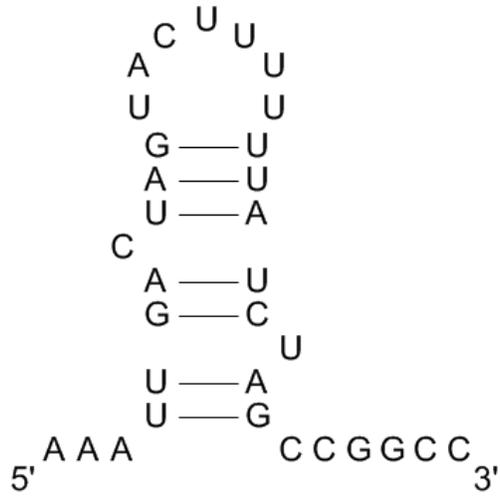

**FIG. 1.** Sequence and structure of a hypothetical pre-miRNA and its dot-bracket notation.

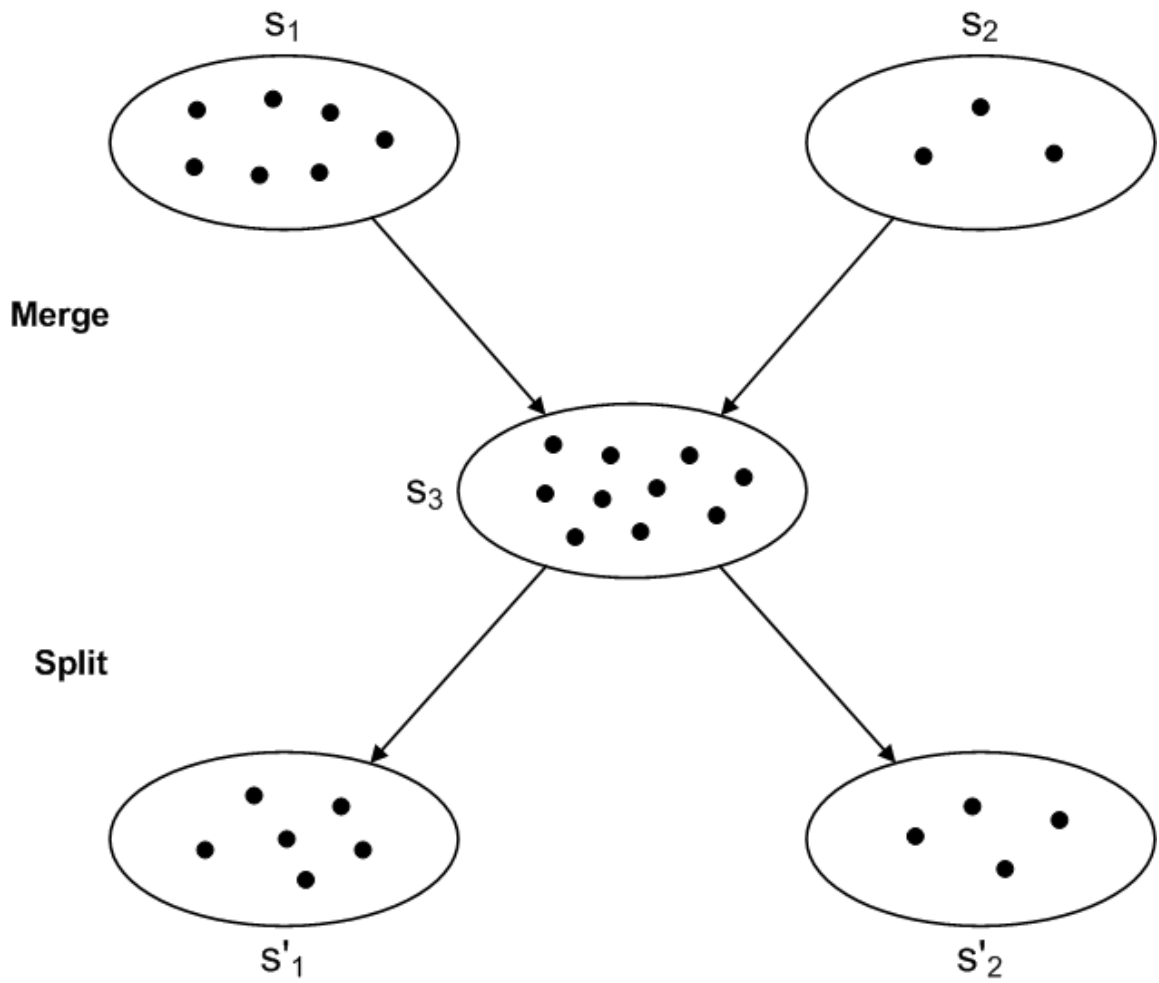

**FIG. 2.** Illustration of the merge and split operations on two feature sets.

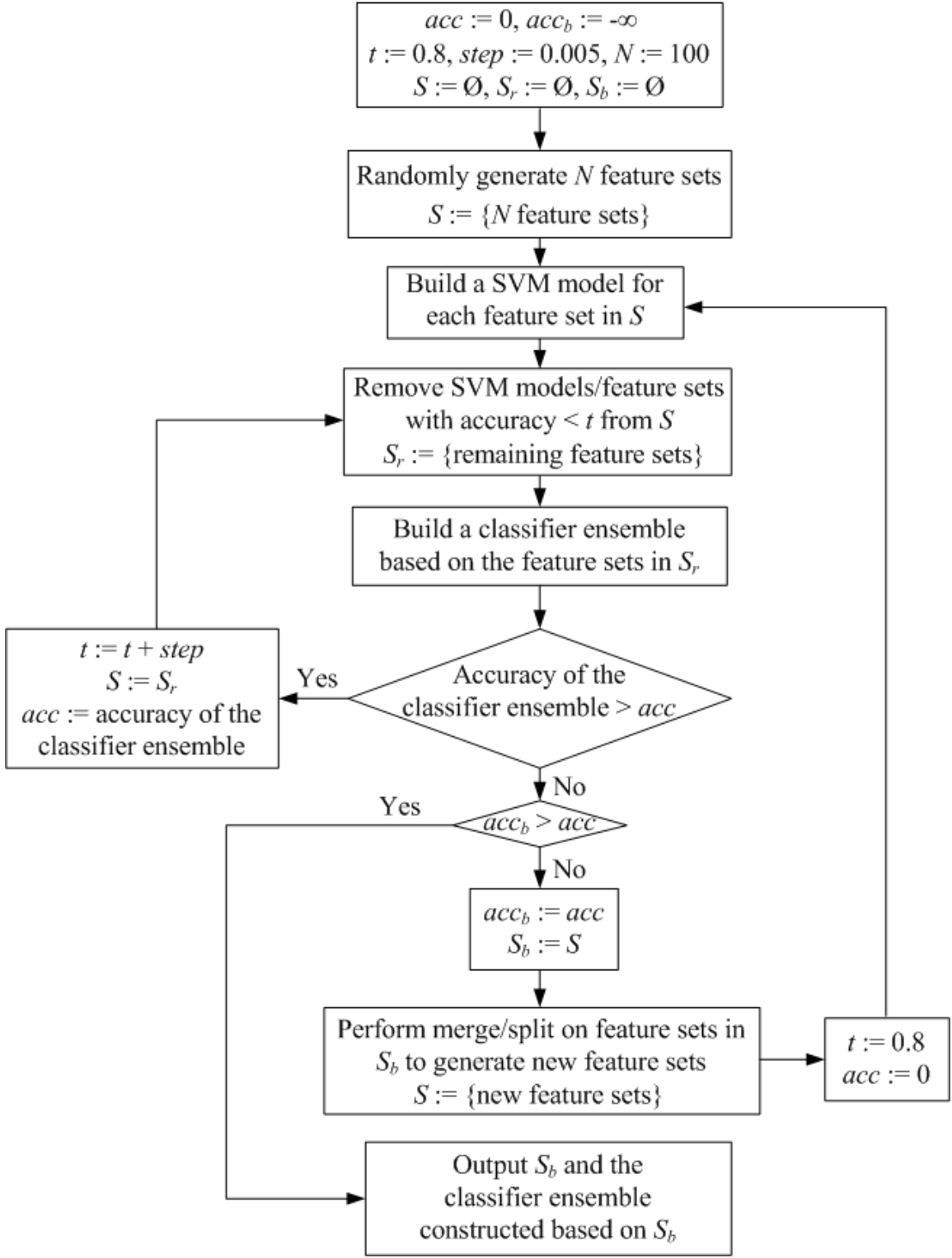

**FIG. 3.** Algorithm for combinatorial feature mining.